\documentclass[10pt,preprint]{aastex}

\usepackage{lineno}

\slugcomment{}

\shorttitle{Discovery of two millisecond pulsars with the Nan\c cay Radio Telescope}
\shortauthors{Cognard et al.}

\begin{document}
\linenumbers


\title{Discovery of two millisecond pulsars in \emph{Fermi} sources with the Nan\c cay Radio Telescope}


\author{
I.~Cognard\altaffilmark{2,1},
L.~Guillemot\altaffilmark{3,1},
T.~J.~Johnson\altaffilmark{4,5,1},
D.~A.~Smith\altaffilmark{6},
C.~Venter\altaffilmark{7},
A.~K.~Harding\altaffilmark{4},
M.~T.~Wolff\altaffilmark{8},
C.~C.~Cheung\altaffilmark{9},
D.~Donato\altaffilmark{5,10},
A.~A.~Abdo\altaffilmark{9},
J.~Ballet\altaffilmark{11},
F.~Camilo\altaffilmark{12},
G.~Desvignes\altaffilmark{13,14},
D.~Dumora\altaffilmark{6},
E.~C.~Ferrara\altaffilmark{4},
P.~C.~C.~Freire\altaffilmark{3},
J.~E.~Grove\altaffilmark{8},
S.~Johnston\altaffilmark{15},
M.~Keith\altaffilmark{15},
M.~Kramer\altaffilmark{3,16},
A.~G.~Lyne\altaffilmark{16},
P.~F.~Michelson\altaffilmark{17},
D.~Parent\altaffilmark{18},
S.~M.~Ransom\altaffilmark{19},
P.~S.~Ray\altaffilmark{8},
R.~W.~Romani\altaffilmark{17},
P.~M.~Saz~Parkinson\altaffilmark{20},
B.~W.~Stappers\altaffilmark{16},
G.~Theureau\altaffilmark{2},
D.~J.~Thompson\altaffilmark{4},
P.~Weltevrede\altaffilmark{16},
K.~S.~Wood\altaffilmark{8}
}

\altaffiltext{1}{Corresponding authors: I.~Cognard, icognard@cnrs-orleans.fr; L.~Guillemot, guillemo@mpifr-bonn.mpg.de; T.~J.~Johnson, tyrel.j.johnson@gmail.com.}
\altaffiltext{2}{Laboratoire de Physique et Chimie de l'Environnement, LPCE UMR 6115 CNRS, F-45071 Orl\'eans Cedex 02, and Station de radioastronomie de Nan\c cay, Observatoire de Paris, CNRS/INSU, F-18330 Nan\c cay, France}
 \altaffiltext{3}{Max-Planck-Institut f\"ur Radioastronomie, Auf dem H\"ugel 69, 53121 Bonn, Germany}
\altaffiltext{4}{NASA Goddard Space Flight Center, Greenbelt, MD 20771, USA}
\altaffiltext{5}{Department of Physics and Department of Astronomy, University of Maryland, College Park, MD 20742, USA}
\altaffiltext{6}{Universit\'e Bordeaux 1, CNRS/IN2p3, Centre d'\'Etudes Nucl\'eaires de Bordeaux Gradignan, 33175 Gradignan, France}
\altaffiltext{7}{North-West University, Potchefstroom Campus, Potchefstroom 2520, South Africa}
\altaffiltext{8}{Space Science Division, Naval Research Laboratory, Washington, DC 20375, USA}
\altaffiltext{9}{National Research Council Research Associate, National Academy of Sciences, Washington, DC 20001, resident at Naval Research Laboratory, Washington, DC 20375, USA}
\altaffiltext{10}{Center for Research and Exploration in Space Science and Technology (CRESST) and NASA Goddard Space Flight Center, Greenbelt, MD 20771, USA}
\altaffiltext{11}{Laboratoire AIM, CEA-IRFU/CNRS/Universit\'e Paris Diderot, Service d'Astrophysique, CEA Saclay, 91191 Gif sur Yvette, France}
\altaffiltext{12}{Columbia Astrophysics Laboratory, Columbia University, New York, NY 10027, USA}
\altaffiltext{13}{Department of Astronomy, University of California, Berkeley, CA 94720-3411, USA}
\altaffiltext{14}{Radio Astronomy Laboratory, University of California, Berkeley, CA 94720, USA}
\altaffiltext{15}{Australia Telescope National Facility, CSIRO, Epping NSW 1710, Australia}
\altaffiltext{16}{Jodrell Bank Centre for Astrophysics, School of Physics and Astronomy, The University of Manchester, M13 9PL, UK}
\altaffiltext{17}{W. W. Hansen Experimental Physics Laboratory, Kavli Institute for Particle Astrophysics and Cosmology, Department of Physics and SLAC National Accelerator Laboratory, Stanford University, Stanford, CA 94305, USA}
\altaffiltext{18}{College of Science, George Mason University, Fairfax, VA 22030, resident at Naval Research Laboratory, Washington, DC 20375, USA}
\altaffiltext{19}{National Radio Astronomy Observatory (NRAO), Charlottesville, VA 22903, USA}
\altaffiltext{20}{Santa Cruz Institute for Particle Physics, Department of Physics and Department of Astronomy and Astrophysics, University of California at Santa Cruz, Santa Cruz, CA 95064, USA}


\begin{abstract}

We report the discovery of two millisecond pulsars in a search for radio pulsations at the positions of \emph{Fermi Large Area Telescope} sources with no previously known counterparts, using the Nan\c cay radio telescope. The two millisecond pulsars, PSRs~J2017+0603 and J2302+4442, have rotational periods of 2.896 and 5.192 ms and are both in binary systems with low-eccentricity orbits and orbital periods of 2.2 and 125.9 days respectively, suggesting long recycling processes. Gamma-ray pulsations were subsequently detected for both objects, indicating that they power the associated \emph{Fermi} sources in which they were found. The gamma-ray light curves and spectral properties are similar to those of previously-detected gamma-ray millisecond pulsars. Detailed modeling of the observed radio and gamma-ray light curves shows that the gamma-ray emission seems to originate at high altitudes in their magnetospheres. Additionally, X-ray observations revealed the presence of an X-ray source at the position of PSR J2302+4442, consistent with thermal emission from a neutron star. These discoveries along with the numerous detections of radio-loud millisecond pulsars in gamma rays suggest that many \emph{Fermi} sources with no known counterpart could be unknown millisecond pulsars. 

\end{abstract}





\section{Introduction}

During its first year of activity, the Large Area Telescope (LAT) aboard the \emph{Fermi Gamma-Ray Space Telescope} \citep{FermiLAT} firmly established millisecond pulsars (MSPs) as bright sources of gamma rays, with the detection of pulsed emission from at least nine Galactic disk MSPs above 0.1 GeV \citep{FermiJ0030,Fermi8MSPs,FermiJ0034}. Normal pulsars were already established as an important class of gamma-ray sources by previous experiments \citep[see e.g.][]{Thompson1999}. The First \emph{Fermi} Catalog of gamma-ray pulsars \citep{FermiPSRCatalog} tabulated the properties of 46 pulsars, including eight millisecond pulsars. In addition, the LAT has observed gamma-ray emission from several globular clusters (GCs) with spectral properties that are consistent with those of populations of MSPs \citep{Fermi47Tuc,FermiGCs} and thus of the flux being due to the combined MSPs in the cluster. 

Millisecond pulsars are rapidly rotating neutron stars (with rotational period of few tens of milliseconds) with very small spin-down rates ($\dot P < 10^{-17}$). They are thought to have acquired their high rotational rate by accretion of matter, and thereby transfer of angular momentum, from a binary companion \citep{Bisnovatyi1974,Alpar1982}, which is now supported by observational evidence \citep{Archibald2009}. About 10\% of the $\sim$ 2000 known pulsars are MSPs, either in the Galactic disk or in globular clusters \citep{ATNFCatalog}. Estimates for the Galactic population of MSPs range from 40000 to 90000 objects \citep[see][and references therein]{Lorimer2008}. A small fraction of these have large enough spin-down luminosities $\dot E$ and small enough distances $d$ to be detectable by the LAT. The minimum $\sqrt{\dot E}/d^2$ of pulsars in the \emph{Fermi} First Pulsar Catalog is 0.1\% of the value for Vela. Furthermore, the sparsity of the photons recorded by the LAT makes MSPs much easier to discover at radio wavelengths than in gamma rays \citep[for a discussion of blind period searches of gamma-ray pulsars, see e.g.][and references therein]{FermiBlindSearch}. However, also when blindly searched in the radio band, the MSPs are difficult targets. On one hand they are faint sources so that their detection generally requires long exposures with large radio telescopes. In addition, most MSPs are in binary systems so the orbital motions need to be taken into account when searching for pulsations, introducing additional parameter combinations, and therefore making data analyses computationally intensive and searches less sensitive than for normal pulsars. 

Radio emission from pulsars is also affected by pulse scattering induced by the ionized component of the interstellar medium, with a characteristic timescale $\tau_s \propto f^{-4} d^2$ where $f$ is the observing frequency and $d$ the pulsar distance \citep{Handbook}. The short rotational periods of MSPs thus introduce an observational bias favoring nearby objects. As a consequence of their proximity and their age, they are more widely distributed in Galactic latitude than normal pulsars. 

The \emph{Fermi} Large Area Telescope First Source Catalog (1FGL) \citep{Fermi1FGL} has 1451 sources, including 630 which are not clearly associated with counterparts known at other wavelengths. The detection of nine radio-loud MSPs in gamma rays strongly suggests that a fraction of high Galactic latitude unassociated \emph{Fermi} sources must be unknown MSPs. Such a source of continuous gamma-ray emission can be deeply scanned for pulsations at radio wavelengths, resulting in MSP discoveries, provided their radio emission beam is pointing toward the Earth. Such searches have been conducted at several radio telescopes around the world, yielding positive results \citep[see e.g.][]{Kerr2011,Keith2011,Ransom2011,Roberts2011}.

Most high Galactic latitude gamma-ray sources are blazars and other Active Galactic Nuclei (AGNs). Fortunately, distinctive indicators of gamma-ray emission from a pulsar are the shape of the spectral emission and the lack of flux variability in gamma rays. Gamma-ray pulsars indeed exhibit sharp cutoffs at a few GeV \citep{FermiPSRCatalog}, while blazars are known to emit above 10 GeV with no sharp energy cutoff (Flat Spectrum Radio Quasars are well-described by broken power-law spectra) \citep{FermiLBAS}. Also, known gamma-ray pulsars are steady sources, whereas blazars show variations of flux over time \citep{Fermi1FGL}. In this exploratory study we limited our source discrimination criterion to spectral shapes. As suggested by \citet{Story2007}, follow-up radio searches of \emph{Fermi} sources having hard spectra with cutoffs should yield discoveries of new MSPs. Gamma-ray variability will be exploited in future studies.

In this article, we present the observations of pulsar candidates made at the Nan\c cay radio telescope that led to the discovery of the MSPs J2017+0603 and J2302+4442 (Section \ref{searchobs}). Following the detections, we made radio timing observations at the Nan\c cay, Jodrell Bank and Green Bank telescopes (see Sections \ref{J2017_timing} and \ref{J2302_timing}). The initial ephemerides for these 2.896 and 5.192 ms pulsars in low-eccentricity orbits around light companions allowed us to detect gamma-ray pulsations in the data recorded by the LAT. In Sections \ref{J2017_gamma}, \ref{J2302_gamma} and \ref{efficiencies} we discuss the gamma-ray properties of the two MSPs, compare the measured light curves and spectral properties with those of previously observed gamma-ray MSPs. We finally present results of radio and gamma-ray light curve modeling in the context of theoretical models of emission in the magnetosphere in Section \ref{modeling}.

\section{Search observations}
\label{searchobs}

The list of 1FGL catalog sources searched for pulsations with the Nan\c cay radio telescope was constructed using the following criteria. The radio search was based on a preliminary list of \emph{Fermi} LAT sources used internally by the instrument team. The selection described here is the same as that used, but was applied to the 1FGL catalog and yielded the same targets. We first removed gamma-ray sources associated with known objects. Sources below $-39^\circ$ in Declination were rejected, as they are not observable with the telescope. Sources with Galactic latitudes $|b| < 3^\circ$ were excluded, being more likely affected by radio pulse scattering and also being less accurately localized in gamma rays because of the intense diffuse gamma-ray background at low Galactic latitudes \citep{Fermi1FGL}. The Nan\c cay beam has a width at half maximum of $4'$ in Right Ascension; therefore we applied a conservative cut by requiring the semi-major axis of the gamma-ray source 95\% confidence ellipse to be less than $3'$. Finally, we selected objects with spectra deviating from simple power laws, \emph{i.e.}, showing evidence for a cutoff, and therefore likely pointing to gamma-ray pulsars. For that we excluded sources with curvature indices below 11.34, the limit at which spectra start departing from simple power laws \citep{Fermi1FGL}. Details on the determination of positions and curvature indices of 1FGL sources can be found in \citet{Fermi1FGL}. 

From these selection criteria we obtained a list of six sources. Four of them, 1FGL~J0614.1$-$3328, J1231.1$-$1410, J1311.7$-$3429 and 1FGL~J1942.7+1033, have been searched for pulsations with the Green Bank and Effelsberg radio telescopes, and radio pulsars have been detected in the first two sources. The results of these searches will be reported elsewhere \citep{Ransom2011,Barr2011}. We carried out radio observations at the Nan\c cay radio telescope of the other two sources in this list, 1FGL~J2017.3+0603 and J2302.8+4443, using the modified Berkeley-Orl\'eans-Nan\c cay (BON) instrumentation \citep{Theureau2005,Cognard2006} at 1.4 GHz. Instead of doing the usual coherent dedispersion of the signal, the code was modified to get a 512 $\times$ 0.25 MHz incoherent filterbank sampled every 32 $\mu$s. The very first data samples were used to determine an amplitude scaling factor and total intensity is recorded as a 4-bit value. Observations were usually one hour long, mainly limited by the fact that Nan\c cay is a meridian telescope.

Data were searched for a periodic dispersed signal using the PRESTO package \citep{presto}. After the standard RFI-excision procedure, a total of 1959 dispersion measure (DM) values up to 1244 pc cm$^{-3}$ were chosen to dedisperse the data. Searches for periodicity were done using the harmonic summing method (up to eight harmonics). We also searched the data for single pulses, and did not find any.

An observation of 1FGL~J2302.8+4443 performed on 2009 November 4 revealed a candidate with a period of 5.192 ms and a DM of 13.4 pc cm$^{-3}$. Confirmation observations scheduled at Nan\c cay and Green Bank (at 350 MHz) later firmly established this new millisecond pulsar. A week after that first discovery, a second candidate in 1FGL~J2017.3+0603 with a period of 2.896 ms and DM of 23.9 pc cm$^{-3}$ was also confirmed with subsequent Nan\c cay and Green Bank Telescope observations as well as with old observations made at the Arecibo telescope. In both cases, substantial variations of the pulsar rotational period were observed, indicating orbital motions, as discussed in Sections \ref{J2017_timing} and \ref{J2302_timing}.

\begin{figure}
\begin{center}
\epsscale{1.}
\plotone{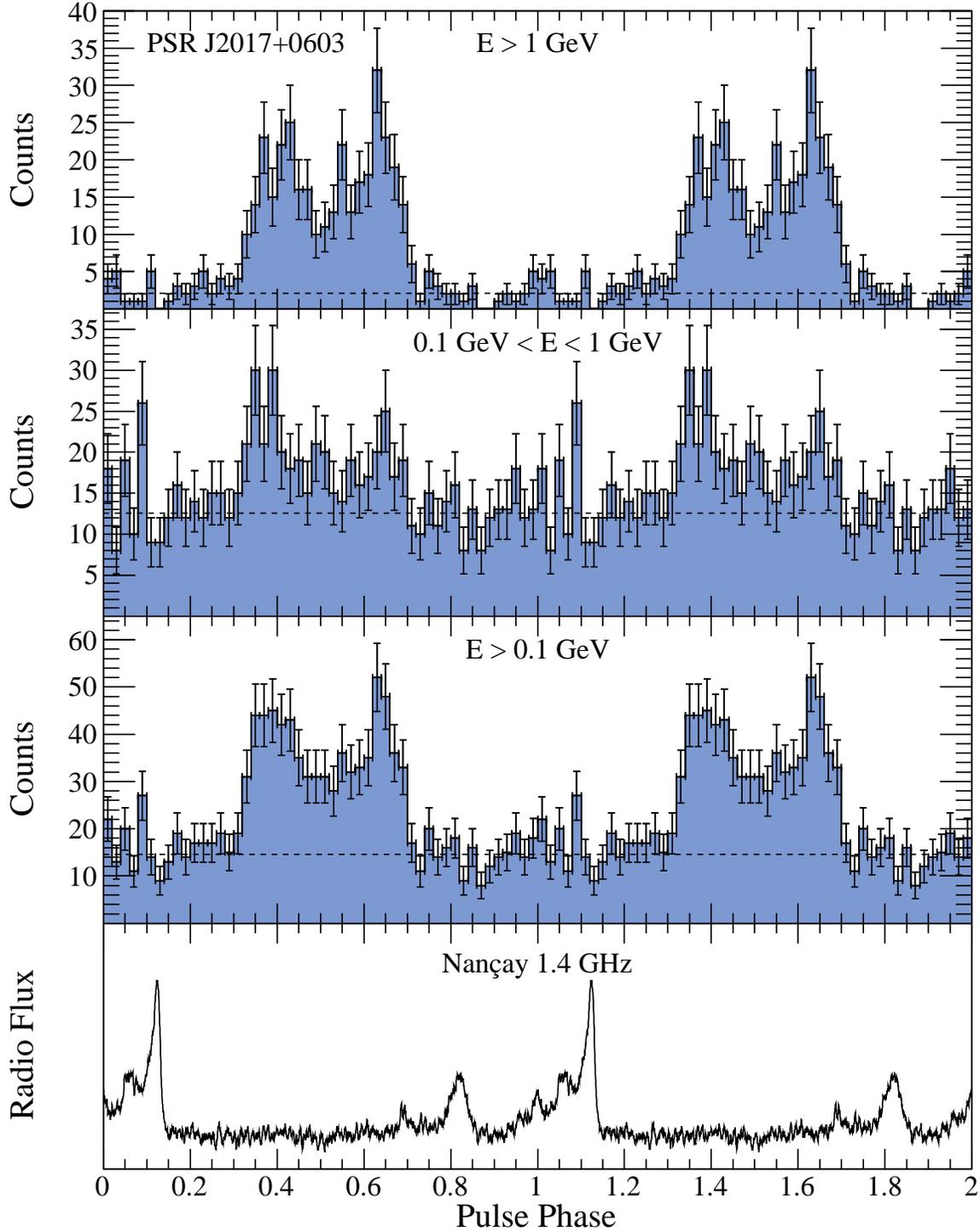}
\caption{Radio and gamma-ray light curves of PSR~J2017+0603. The bottom panel shows an integrated radio profile at 1.4 GHz with 2048 bins per rotation, recorded with the Nan\c cay radio telescope, based on 16.2 hours of coherently dedispersed observations. The top three panels show light curves in different energy bands (labeled) for gamma-ray events within 0.8$^\circ$ of the pulsar position, with 50 bins per rotation. Two full rotations are shown for clarity. See Section \ref{J2017_gamma} for details on the determination of background levels, shown by horizontal dashed lines.\label{phaso_J2017}}
\end{center}
\end{figure}

\begin{figure}
\begin{center}
\epsscale{1.}
\plotone{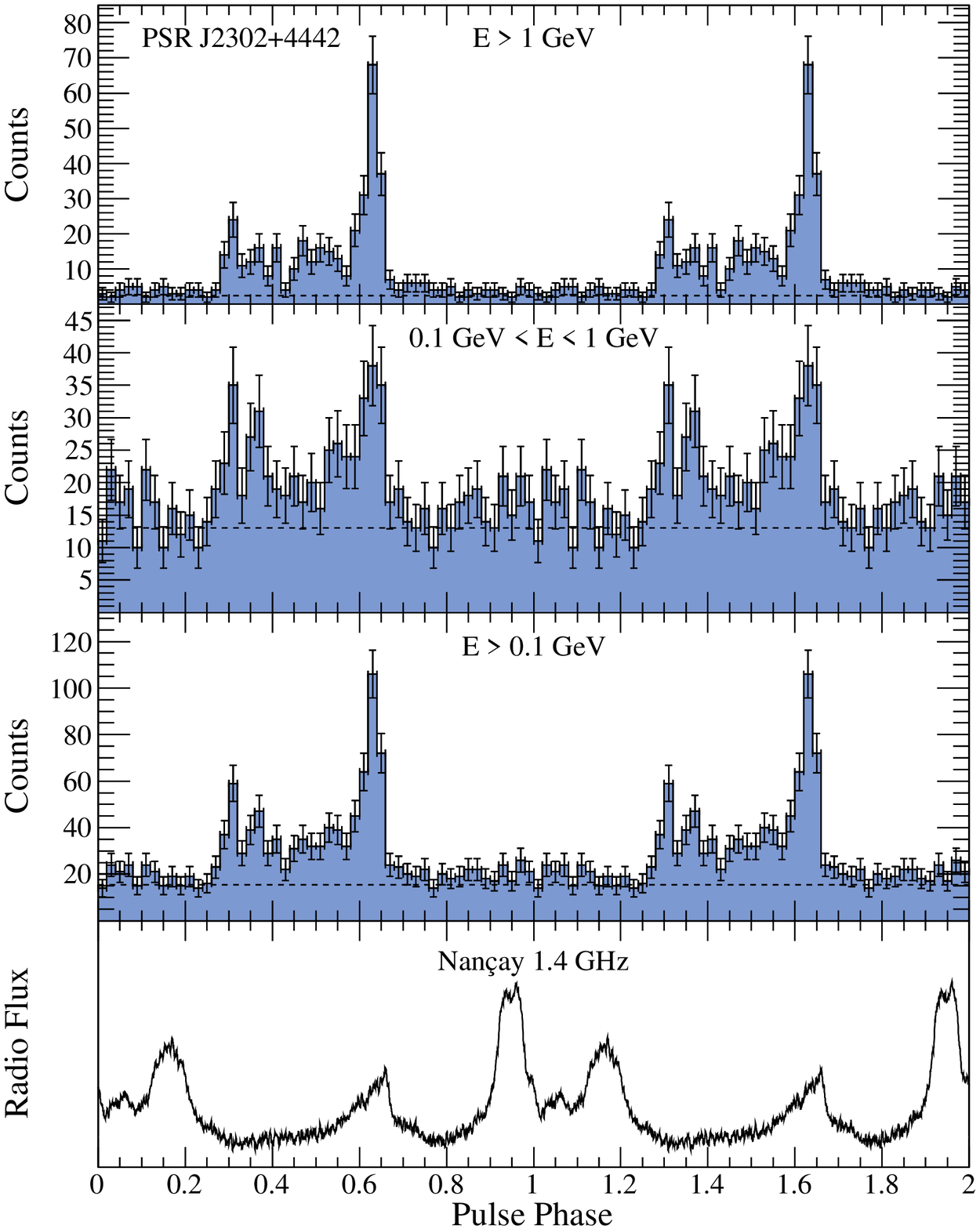}
\caption{Same as Figure \ref{phaso_J2017}, for PSR~J2302+4442. The radio profile is based on 20.9 hours of observation.\label{phaso_J2302}}
\end{center}
\end{figure}

Integrated radio profiles at 1.4 GHz are presented in Figures \ref{phaso_J2017} and \ref{phaso_J2302}. The pulse profile of PSR~J2017+0603 is complex and exhibits at least five components. A sharp peak is observed, making PSR~J2017+0603 a promising addition to pulsar timing array programs. The radio profile of PSR~J2302+4442 is broad, with at least four pulsed components, three of which form a first structure whose mid-point is separated by $\sim$ 0.6 rotation from the fourth component. The mean flux density averaged over all observations for the two pulsars was determined using a calibrated pulse noise diode fired for 10 seconds before each observation \citep[see][for a description of radio flux measurements with the Nan\c cay radio telescope]{Theureau2011}. PSR~J2017+0603 presents a mean flux density at 1.4 GHz of 0.5 $\pm$ 0.2 mJy, while PSR~J2302+4442 is brighter at 1.2 $\pm$ 0.4 mJy, both being typical values for millisecond pulsars.

\section{PSR~J2017+0603}

\subsection{Timing observations}
\label{J2017_timing}

After the initial discovery of PSR~J2017+0603, timing observations were undertaken at the Nan\c cay radio telescope and the Lovell telescope at the Jodrell Bank Observatory \citep{Hobbs2004}. Nan\c cay timing observations were done using two different configurations of the BON instrumentation described above. Between MJDs 55142 and 55228 we used the 512 $\times$ 0.25 MHz incoherent filterbank at 1334 MHz, and the standard coherent dedispersor \citep{cognard2009} between MJDs 55232 and 55342. The coherent dedispersion is performed in 4 MHz channels over a total bandwidth of 128 MHz centered at 1408 MHz. Eighteen times of arrival (TOAs) were recorded with the filterbank BON with a mean uncertainty of the TOA determination of 6.3 $\mu$s, and 19 TOAs were measured with the coherent dedispersor BON with a mean uncertainty of 2.5 $\mu$s. In addition, 24 radio TOAs were recorded with the Lovell telescope at 1520 MHz between MJDs 55218 and 55305, with a mean uncertainty of 17.8 $\mu$s. These data were used to derive an initial timing solution covering the first seven months post-discovery, using the TEMPO2 pulsar timing package\footnote{http://sourceforge.net/projects/tempo2/} \citep{tempo2}. The dispersion measure was estimated independently: the data recorded with the BON backend of the Nan\c cay telescope were cut in four frequency bands of 32 MHz, centered at 1358, 1390, 1422 and 1454 MHz. We fitted the multi-frequency dataset with the initial timing solution, where the DM was left free. We measured DM $= 23.918 \pm 0.003$ pc cm$^{-3}$. 

For this DM and line-of-sight, the NE2001 model of the Galactic distribution of free electrons\footnote{Available at http://rsd-www.nrl.navy.mil/7213/lazio/ne\_model/} assigns a distance of 1.56 $\pm$ 0.16 kpc \citep{NE2001}. Archival optical and infrared images (POSS-II) and radio images (NVSS) show no obvious clouds which might indicate electron overdensities. The line-of-sight intersects the Galaxy's Sagittarius arm at about 2 kpc from the Earth \citep{Reid2009} and at the nominal DM distance the pulsar environment is not especially crowded. Nevertheless, density variances not modeled in NE2001 could change the distance significantly.

We phase-folded the data recorded by the \emph{Fermi} LAT using the initial timing solution, and detected pulsed gamma-ray emission with high significance. The gamma-ray light curve and spectral properties of the MSP are discussed below. However, we observed gradual phase coherence loss for gamma-ray photon dates which were earlier than the ephemeris validity interval, defined by the radio observation time span, indicating erroneous parameters in the initial timing solution. To enhance the timing solution and make it accurate for the entire time range of the LAT data used here, we extracted TOAs from the gamma-ray data using the method described in \citet{Ray2011}. The LAT data were divided in time intervals where the gamma-ray pulsation had a significance of at least 3$\sigma$. For each time interval, we then measured a TOA by cross-correlating the observed gamma-ray light curve and a standard template, derived from the fraction of the LAT data covered by the timing solution. The pulsar ephemeris was then optimized with the gamma-ray and radio TOAs. This procedure was repeated until phase-coherence was ensured over the whole LAT dataset. We eventually extracted a total of 10 gamma-ray TOAs between MJDs 54682 and 55294, with a mean uncertainty of 49.1 $\mu$s. 

The final timing solution was built using radio and gamma-ray TOAs, fitting for the pulsar position, rotational period and first derivative, binary parameters and phase jumps between observatories. The dispersion measure value was held fixed at this stage. The low-eccentricity orbit was described using the ELL1 model \citep{Lange2001}. We corrected for any underestimation of TOA uncertainties and badness of fit by using ``error factors'' (parameters EFAC in TEMPO2) on each set of TOAs, following the method described in \citet{Verbiest2009}, in order to get a reduced $\chi^2$ value as close as possible to unity for the entire dataset. We obtained a reduced $\chi^2$ value of 1.14. The corresponding timing solution is given in Table \ref{ephem}. The spin-down luminosity and magnetic field at the light cylinder derived from the measured period and period derivative are typical of other gamma-ray MSPs detected so far \citep{Fermi8MSPs,FermiJ0034}. However, with a small period derivative of $\simeq 8.3 \times 10^{-21}$ and at a distance of 1.56 kpc according to the NE2001 model, PSR~J2017+0603 is subject to significant contribution from the Shklovskii effect \citep{Shklovskii1970}, making the apparent period derivative greater than the intrinsic one, by $2.43 \times 10^{-21} \mathrm{s}^{-1} P d \mu_T^2$, where $P$ is the pulsar rotational period in s, $d$ is the distance in kpc, and $\mu_T$ is the proper motion, in mas yr$^{-1}$. This effect would reduce the true $\dot P$ and thus reduce the calculated spin-down luminosity and magnetic field at the light cylinder. In this study we could not measure any significant proper motion, though it may become possible with accumulated radio observations. 

Using the measured binary parameters, projected semi-major axis of the orbit,  $x$, and orbital period, $P_b$, we calculated the mass function in Table \ref{ephem}, given by $f(m_p,m_c) = (m_c \sin i)^3 / (m_p + m_c)^2 = (4 \pi^2 c^3 x^3) / (G \mathrm{M}_\Sun P_b^2)$ where $m_p$ is the pulsar mass, $m_c$ is the companion mass and $i$ is the inclination of the orbit. Assuming an edge-on orbit ($i = 90^\circ$) and a pulsar mass of 1.4 M$_\Sun$, we calculate a lower limit on $m_c$ of 0.18 M$_\Sun$. As noted in \citet{Handbook}, the probability of observing a binary system with an inclination of less than $i_0$ for a random distribution of orbital inclinations is $1 - \cos(i_0)$, therefore a 90\% confidence upper limit on the companion mass can be derived by assuming an inclination angle $i$ of 26$^\circ$. Doing so gives an upper limit of 0.45 M$_\Sun$ for the companion mass of PSR~J2017+0603. These mass function and range of likely companion mass values indicate that the companion star probably is a \emph{He}-type white dwarf. 

\subsection{Optical, UV and X-ray analysis}
\label{J2017_X}

We searched for X-ray and optical/UV counterparts in \emph{Swift} \citep{Gehrels2004} observations obtained from Feb.-Mar.~2009. In an XRT \citep{Burrows2005} image with 16.4 ks of cumulative exposure, we measured an upper limit to the 0.5 -- 8 keV count rate of $< 1.5$ counts ks$^{-1}$ at the position of PSR~J2017+0603. Adopting a flux conversion of $5 \times 10^{-11}$ erg cm$^{-2}$ counts$^{-1}$ (0.3 -- 10 keV) from \citet{Evans2007}, and an appropriate conversion to our choice of energy range, results in a flux limit between 0.5 and 8 keV of $< 6 \times 10^{-14}$ erg cm$^{-2}$ s$^{-1}$. The UVOT \citep{Roming2005} images show a relatively bright field source \citep[B = 19.8 mag, R.A. = 20:17:22.51, Decl. = +06:03:07.7 with $<$ $0.1''$ uncertainty, from][]{Monet2003}, that is $3.6''$ away from the pulsar position, which contaminates the photometry. Moving the aperture sufficiently to avoid this source, we estimate optical/UV upper limits for the pulsar to be 80 (V), 47 (B), 17 (U), 7 (W1), 5 (M2), and 3 (W2) $\mu$Jy. All flux upper limits are at the 3$\sigma$ confidence level.

\subsection{Gamma-ray analysis}
\label{J2017_gamma}

The gamma-ray data recorded by the LAT were analyzed using the \emph{Fermi} science tools (STs) v9r16p1\footnote{http://fermi.gsfc.nasa.gov/ssc/data/analysis/scitools/overview.html}. Using \emph{gtselect} we selected events recorded between 2008 August 4 and 2010 May 26, with energies above 0.1 GeV, zenith angles $\leq$ 105$^\circ$, and within 20$^\circ$ of the pulsar's position. We furthermore selected events belonging to the ``Diffuse'' class of events under the P6\_V3 instrument response function (IRFs), those events having the highest probability of being photons \citep{FermiLAT}. We finally rejected times when the rocking angle of the satellite exceeded 52$^\circ$, required that the DATA\_QUAL and LAT\_CONFIG are equal to 1 and that the Earth's limb did not infringe upon the Region of Interest (ROI) using \emph{gtmktime}. Finally, we phase-folded gamma-ray events using the pulsar ephemeris given in Table \ref{ephem} and the \emph{Fermi} plug-in now distributed with the TEMPO2 pulsar timing package.

Figure \ref{phaso_J2017} shows radio and gamma-ray light curves of PSR~J2017+0603, for gamma-ray events within 0.8$^\circ$ of the pulsar. Under this cut, most high-energy photons (energies above 1 GeV) coming from the pulsar are kept, while the contribution of background emission, mostly present at lower energies, is reduced. The bin-independent \emph{H}-test parameter \citep{deJager1989,deJager2010} has a value of 235, corresponding to a pulsation significance well above 10$\sigma$. As can be seen in Figure \ref{phaso_J2017}, the gamma-ray pulse profile comprises two close peaks, offset from the radio emission. The absolute phasing in these light curves is such that the maximum of the first Fourier harmonic of the radio profile transferred back into the time domain is at phase 0. Under that convention, the maximum of the radio profile is at $\Phi_r = 0.123$ in phase. We fitted the gamma-ray light curve above 0.1 GeV using a two-sided Lorentzian function for the asymetrical first peak and a simple Lorentzian function for the second peak above constant background. For each peak, the peak position $\Phi_i$ and the Full Width at Half-Maximum FWHM$_i$ are listed in Table \ref{params}. The Table also lists the values of the radio-to-gamma-ray lag $\delta = \Phi_1 - \Phi_r$, and the gamma-ray peak separation $\Delta = \Phi_2 - \Phi_1$. Quoted uncertainties are statistical. For the radio-to-gamma-ray lag $\delta$ we quote a second error bar, reflecting the uncertainty on the conversion of a TOA recorded at 1.4 GHz to infinite frequency, due to the uncertainty on the dispersion measure (DM) value given in Table \ref{ephem}. With $\delta \simeq 0.22$ and $\Delta \simeq 0.29$, PSR~J2017+0603 follows the correlation between $\delta$ and $\Delta$ expected in outer magnetospheric models as pointed out by \citet{Romani1995} and effectively observed for currently known gamma-ray pulsars \citep[see Figure 4 of][]{FermiPSRCatalog}. However it is interesting to note that this MSP occupies a region of the $\delta$ -- $\Delta$ plot where few gamma-ray pulsars were known.  

The spectral analysis was done by fitting the region around PSR~J2017+0603 using a binned likelihood method \citep{Cash1979,Mattox1996}, implemented in the \emph{pyLikelihood} module of the \emph{Fermi} STs. All 1FGL catalog sources \citep{Fermi1FGL} within 15$^\circ$ from the pulsar as well as additional point sources found in an internal LAT source list using 18 months of data were included in the model. Sources were modeled with power-law spectra, except for PSR~J2017+0603 which was modeled with an exponentially cut off power-law, of the form:

\begin{eqnarray}
\frac{dN}{dE} = N_0 \left( \frac{E}{1 \mathrm{GeV}} \right)^{-\Gamma} \exp \left[ - \left( \frac{E}{E_c}\right)^\beta \right].
\label{model}
\end{eqnarray}

\begin{figure}
\begin{center}
\epsscale{1.}
\plotone{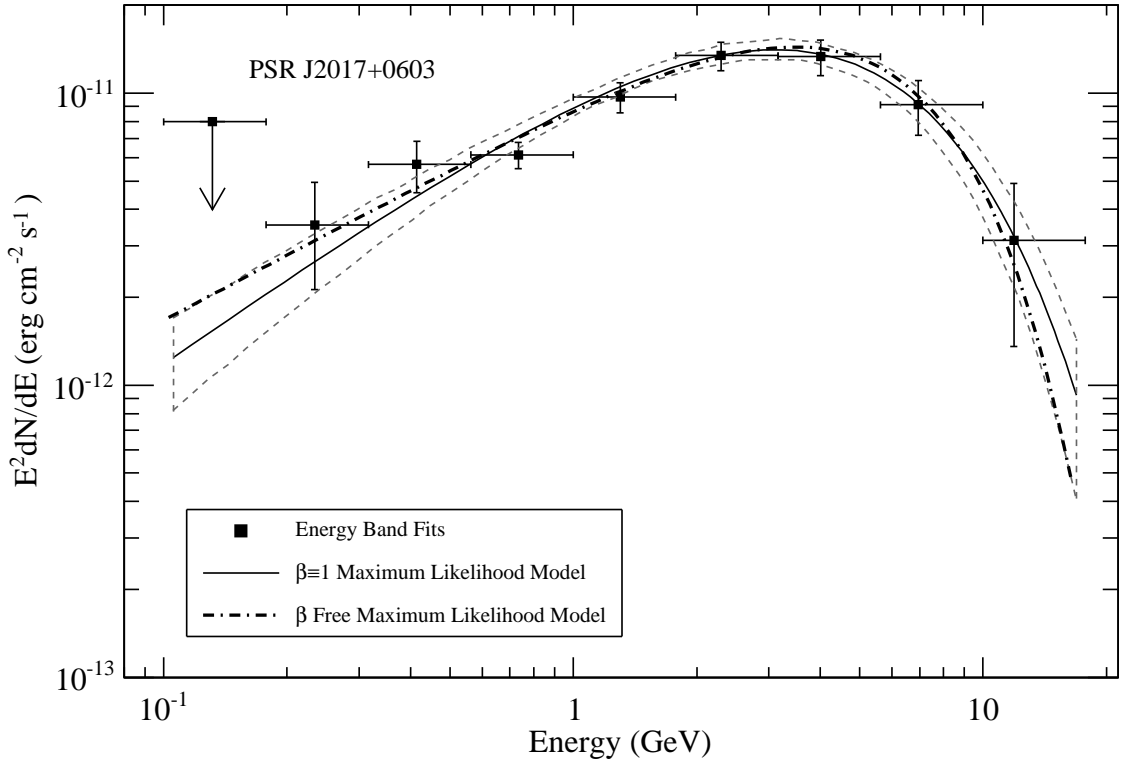}
\caption{Phase-averaged gamma-ray energy spectrum for PSR~J2017+0603. The solid black line shows the best-fit model from fitting the full energy range with a simple exponentially cutoff power-law functional form ($\beta \equiv 1$). Dashed lines indicate 1$\sigma$ errors on the latter model. The dot-dashed line represents the spectral fit with the $\beta$ parameter left free. Data points are derived from likelihood fits of individual energy bands where the pulsar is modeled with a simple power-law form. A 95\% confidence level upper limit was calculated for any energy band in which the pulsar was not detected above the background with a significance of at least 2$\sigma$. \label{spectre_J2017}}
\end{center}
\end{figure}

In Equation (\ref{model}), $N_0$ is a normalization factor, $\Gamma$ denotes the photon index, and $E_c$ is the cutoff energy of the pulsar spectrum. The parameter $\beta$ determines the steepness of the exponential cutoff. \emph{Fermi} LAT pulsar spectra are generally well-described by a simple exponential model, $\beta \equiv 1$. The Galactic diffuse emission was modeled using the \emph{gll\_iem\_v02} mapcube, while the extragalactic diffuse and residual instrument background components were modeled using the \emph{isotropic\_iem\_v02} template\footnote{The diffuse models are available through the \emph{Fermi Science Support Center} (FSSC) (see http://fermi.gsfc.nasa.gov/ssc/)}. Normalization factors and indices for all point sources within 7$^\circ$ from PSR~J2017+0603 and normalization factors for diffuse components were left free. The best-fit values for the photon index and cutoff energy of PSR~J2017+0603 for a simple exponentially cut off power-law ($\beta = 1$) are listed in Table \ref{params}, and the corresponding gamma-ray energy spectrum is shown in Figure \ref{spectre_J2017}. The first errors are statistical, and the second are systematic. These last uncertainties were calculated by following the same procedure as above, but using bracketing IRFs for which the effective area has been perturbed by $\pm$ 10\% at 0.1 GeV, $\pm$ 5\% near 0.5 GeV and $\pm$ 20\% at 10 GeV with linear interpolations in log space between. We also modeled the millisecond pulsar with a power-law fit, $\beta = 0$, and found that the exponentially cut off power-law model ($\beta = 1$) is preferred at the 9$\sigma$ level. A fit of the pulsar's spectrum with the $\beta$ parameter in Equation (\ref{model}) left free led to $\beta = 1.5 \pm 0.6$. This value is consistent with 1 within statistical errors, and the extra free parameter did not improve the quality of the fit, as can be seen in Figure \ref{spectre_J2017}. We therefore conclude that the simple exponentially cut off power-law model (with $\beta = 1$) reproduces the present data well. 

With the full spectral model obtained with this analysis and the \emph{Fermi} ST \emph{gtsrcprob}, we calculated probabilities that each photon originates from the different gamma-ray sources in the ROI. If we denote $\omega_i$ as the probability that a given photon has been emitted by PSR~J2017+0603, and therefore $(1 - \omega_i)$ the probability that the photon is due to background, then the background level in the considered ROI can be estimated by calculating $b = \sum_{i}^{N} (1 - \omega_i)$, where $N$ is the number of photons in the ROI. The background levels shown in Figure \ref{phaso_J2017} were calculated with this method, which is more powerful at discriminating background events than methods involving surrounding annuli.

The photon index $\Gamma$ and cutoff energy $E_c$ measured in this analysis are reminiscent of those of previously-detected gamma-ray MSPs \citep{Fermi8MSPs,FermiJ0034}. Integrating Equation (\ref{model}) above 0.1 GeV yields the photon flux $F$ and energy flux $G$ given in Table \ref{params}. The 1FGL Catalog quotes an energy flux above 0.1 GeV for 1FGL~J2017.3+0603 of (4.5 $\pm$ 0.5) $\times 10^{-11}$ erg cm$^{-2}$ s$^{-1}$, consistent with the value measured for PSR~J2017+0603. Nevertheless, the high-redshift blazar CLASS~J2017+0603 \citep{Myers2003,Fermi1LAC} located $2.3'$ from the pulsar could also contribute to the gamma-ray flux of the 1FGL source. We checked that hypothesis by selecting the off-peak region of the spectrum (pulse phases between 0.25 and 0.75) and by performing a likelihood analysis of the selected data, where the blazar was modeled by a power-law. Following this procedure we did not detect any significant emission from the blazar. PSR~J2017+0603 therefore is the natural counterpart of 1FGL~J2017.3+0603. 

\section{PSR~J2302+4442}

\subsection{Timing observations}
\label{J2302_timing}

Radio timing observations of the pulsar in 1FGL~J2302.8+4443 were conducted at the Nan\c cay radio telescope in the two configurations described in Section \ref{J2017_timing}, the Green Bank Telescope in West Virginia with the GUPPI backend\footnote{https://safe.nrao.edu/wiki/bin/view/CICADA/NGNPP}, and the Lovell telescope at the Jodrell Bank Observatory. Between MJDs 55139 and 55218, 29 TOAs were recorded with the filterbank BON with a mean uncertainty on the determination of arrival times of 7.6 $\mu$s, while the coherent dedispersor was used to measure 22 TOAs between MJDs 55150 and 55342, with a mean uncertainty of 2.1 $\mu$s. The Green Bank Telescope recorded 32 TOAs in two observation sessions, at MJDs 55095 and 55157, with a mean uncertainty of 5.1 $\mu$s. The Lovell telescope recorded a total of 38 TOAs at 1520 MHz between MJDs 55217 and 55304, with a mean uncertainty of 20.4 $\mu$s. An initial timing solution was built using these radio timing observations and the TEMPO2 pulsar timing package. As with PSR~J2017+0603, data recorded with the BON backend were cut in four frequency bands of 32 MHz, and the multi-frequency TOAs extracted from these observations were used to determine the dispersion measure. 

We measured DM $= 13.762 \pm 0.006$ pc cm$^{-3}$. The NE2001 model assigns this DM and line-of-sight a distance of 1.18$^{+0.10}_{-0.23}$ kpc. Again, optical, infrared, and radio images show no clouds. These line-of-sight and distance place the pulsar within the Orion spur of the Sagittarius arm. As above, unmodeled electron density variations could change the distance significantly.

We used the initial timing solution to phase-fold the LAT data and detected highly-significant gamma-ray pulsations. The gamma-ray light curve and spectral properties of the MSP are discussed below. Similarly to PSR~J2017+0603, we could not fold all LAT data properly using the initial timing solution, as we observed loss of phase-coherence for photons recorded before the first radio timing data were taken. Following the iterative procedure described in Section \ref{J2017_timing}, we extracted TOAs for the gamma-ray data, optimized the timing solution by adding the gamma-ray TOAs to the radio dataset, and phase-folded the LAT data until we obtained phase-coherence over the entire \emph{Fermi} dataset described previously. We finally measured nine TOAs between MJDs 54682 and 55294 with an uncertainty of 44.6 $\mu$s. 

The final timing solution obtained by fitting for the pulsar position, rotational period and first time derivative and binary parameters is listed in Table \ref{ephem}. The low-eccentricity orbit of PSR~J2302+4442 was also described using the ELL1 model. The same procedure to correct underestimated TOA uncertainties with EFAC parameters as described in \ref{J2017_timing} was used, resulting in a reduced $\chi^2$ value of 1.04. Like PSR~J2017+0603, J2302+4442 is subject to significant contribution from the Shklovskii effect, with a relatively small period derivative of $\simeq 1.33 \times 10^{-20}$. We were not able to measure any significant proper motion with the present dataset, however accumulated radio observations may help constrain the Shklovskii contribution. 

Under the assumption of an edge-on orbit and a pulsar mass of 1.4 M$_\Sun$, the lower limit on the companion mass is found to be 0.30 M$_\Sun$. However, assuming an inclination of $i = 26^\circ$ leads to an upper limit of 0.81 $M_\Sun$ for the companion mass, suggesting that the companion star could either be a \emph{He}-type or a \emph{CO}-type white dwarf. Nevertheless, the orbital period and eccentricity of PSR~J2302+4442 are in good agreement with the $P_b$ -- $e$ relationship predicted by \citet{Phinney1992}, whereas ``intermediate-mass binary pulsars'' (IMBPs) with heavier companion stars do not necessarily follow the relationship. This suggests that PSR~J2302+4442 is in orbit with a low-mass \emph{He}-type companion, and thus that its inclination angle $i$ must be large. Future radio timing observations may help determine the companion mass and orbital inclination, via the measurement of the Shapiro delay \citep[see e.g.][]{Handbook}. As discussed in detail in \citet{Freire2010}, the amplitude of the measurable part of the Shapiro delay for an orbit with medium to high inclination is proportional to $h_3 = T_\Sun m_c \times \left(\sin(i) / (1 + |\cos(i)|)\right)$, where $T_\Sun = G M_\Sun / c^3 \sim 4.925\ 490\ 947$ $\mu$s. With a current average uncertainty on TOAs recorded with the Nan\c cay BON backend of $\sim$ 2.1 $\mu$s, we expect the Shapiro delay to be measurable for large $m_c$ and $i$ values.

\subsection{Optical, UV and X-ray analysis}

In the \emph{Swift} XRT image of the PSR~J2302+4442 field (9.1 ks summed exposure), there is a marginal detection (2.6$\sigma$) of an X-ray source (R.A. = 23:02:47.00, Decl. = +44:42:20.7; 90\% confidence radius of $6.3''$) that is consistent with the pulsar position. The 0.5 -- 8 keV flux corresponding to the observed count rate of (1.0 $\pm$ 0.4) counts ks$^{-1}$ is $\sim 4 \times$10$^{-14}$ erg cm$^{-2}$ s$^{-1}$ (See Section \ref{J2017_X} for details on the flux conversion). The optical and UV upper limits at the pulsar position are: 53 (V), 26 (B), 13 (U), 6 (W1), 4 (M2), and 3 (W2) $\mu$Jy.

\begin{figure}
\begin{center}
\epsscale{1.}
\plotone{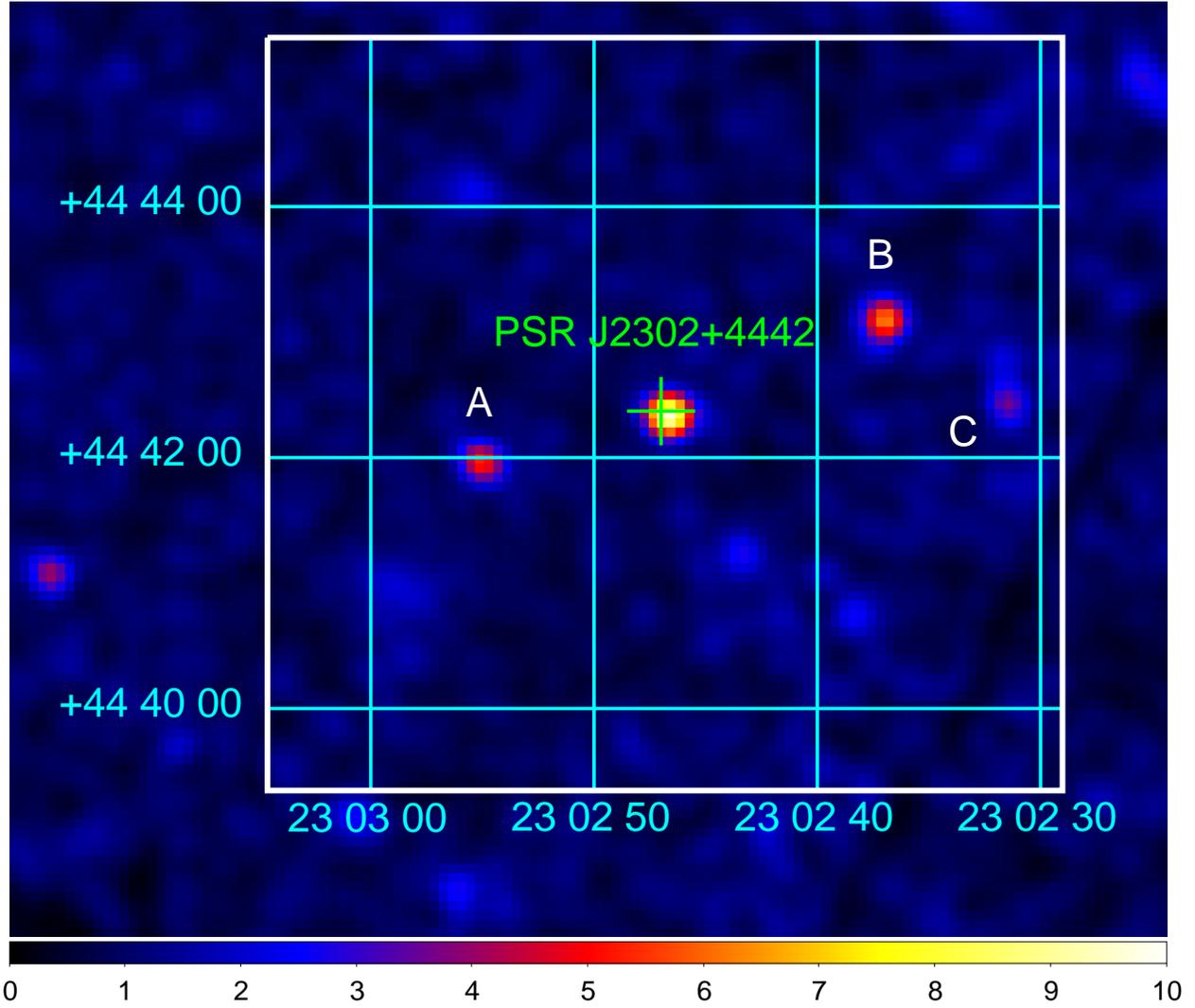}
\caption{The combined EPIC-MOS1 and -MOS2 image of the field of the pulsar PSR J2302+4442, based on 24.9 and 25.1 ks exposures, respectively, and smoothed by 3 pixel widths ($\sim 3.3''$). The color scale represents counts per pixel. The position of the pulsar is shown by the green cross and is indistinguishable to the accuracy of the X-ray image from the position of the X-ray source we call XMMUJ230247+444219. This source, and the source labeled A, were both detected by the \emph{Swift} XRT in its exploration of this field (see text) but the two other labeled sources (B \& C) apparently were not detected by the XRT.\label{XMM_map}}
\end{center}
\end{figure}

On 2009 December 25, while this \emph{Fermi} LAT source was as yet unidentified, the XMM-Newton satellite observed the LAT-source field with the EPIC-MOS and -PN cameras in an effort to explore the source region. We reduced these data with the Science Analysis Software (SAS) version 10.0.0 released on 2010 April 28. After filtering the observation for intervals of high particle background we were left with good time intervals consisting of 24.9 ks, 25.1 ks, and 20.8 ks exposures in the EPIC-MOS1, -MOS2, and -PN instruments, respectively. A number of sources were detected in the field of the \emph{Fermi} LAT source, as can be seen in Figure \ref{XMM_map}. Once the radio pulsar position was refined to the arcsecond level one X-ray source in particular was positionally identified as the likely pulsar candidate and we name this source \mbox{XMMUJ230247+444219}.

We extracted events from a $50''$ region around the source from both the MOS1 and MOS2 event files, and background events from a $100''$ region nearby and apparently free of faint X-ray sources but still on the same respective MOS CCD chips. For the PN event files, in order to avoid a gap between adjacent CCDs, we extracted events from a region only $10''$ in radius and a background region of radius $80''$. From the MOS instruments we obtain 269 and 262 events, and from the PN we obtain 176 events, respectively, from the source regions. This yields, along with the background estimates, a combined detection significance of 13.2$\sigma$ from all three detectors for XMMUJ230247+444219.

We grouped these events into spectral bins of at least 20 counts per bin and performed a simultaneous XSPEC\footnote{http://heasarc.gsfc.nasa.gov/docs/xanadu/xspec/} fit to an absorbed power-law model to all three spectra in the 0.4 to 3.0 keV range. This yields a power-law index of 5.9 which we regard as unphysical and so we discard this model. On the other hand, an absorbed neutron star hydrogen atmosphere model \citep[phabs $\times$ nsatmos, see][]{Heinke2006} yields an acceptable fit, provided that the neutron star mass and radius are fixed at 1.4 M$_\Sun$ and 10.0 km, respectively, and the source distance is fixed at the DM value of 1.18 kpc. However, while we obtain an acceptable reduced $\chi^2$ of 1.032 for 17 degrees of freedom, we measure a column density of N$_\mathrm{H}$ = 0.018$^{+0.31}_{-0.018} \times 10^{22}$ cm$^{-2}$ (90\% confidence) meaning that N$_\mathrm{H}$ is poorly constrained and consistent with values anywhere from zero to greater than $3 \times 10^{21}$ cm$^{-2}$. Also, the temperature range is T$_\mathrm{eff}$ = $1.2^{+0.4}_{-0.7} \times 10^6$ K (90\% confidence) where T$_\mathrm{eff}$ is observed at infinity. The large error ranges for T$_\mathrm{eff}$ and N$_\mathrm{H}$ prompt us to try to reduce parameter uncertainties by better constraining $N_\mathrm{H}$, within the context of this same atmospheric emission model, by considering an independent analysis of the same direction.

The value of N$_\mathrm{H}$ obtained from the LAB Survey of Galactic HI \citep{Kalberla2005} for this direction in the Galaxy, 1.32 $\times 10^{21}$ cm$^{-2}$, is well within the wide range of acceptable column densities obtained in the above model fit. If we now fix N$_\mathrm{H}$ at this value in the same absorbed neutron star atmospheric model we obtain a new fit with reduced $\chi^2$  of 1.000 with 18 degrees of freedom and more precise error ranges: T$_\mathrm{eff} = 8.1^{+1.8}_{-1.4} \times 10^5$ K (90\% confidence) where T$_\mathrm{eff}$ is again observed at infinity. The model normalization is $1.81^{+3.06}_{-1.16} \times 10^{-2}$ (90\% confidence). The derived unabsorbed X-ray flux in the 0.5 to 3 keV range is $3.1^{+0.4}_{-0.4} \times 10^{-14}$ erg cm$^{-2}$ s$^{-1}$ (90\% confidence). The model normalization gives an indication of the fraction of the neutron star surface that is emitting and amounts to a total of $\simeq$ 23 km$^2$ in our simple model, less than the entire neutron star surface area. We note that after accounting for the observed background, we are working with approximately 300 observed source counts and given this small number and the restricted energy range we cannot set strong limits on the column density to the source nor can we investigate the possibility of a non-thermal component above 2 keV in the X-ray spectrum. Thus, while it is very likely that this X-ray source is in fact the pulsar PSR~J2302+4442, longer duration X-ray observations with XMM-Newton or \emph{Chandra} are required to more precisely determine its atmospheric parameters and search for possible X-ray pulsations.

\subsection{Gamma-ray analysis}
\label{J2302_gamma}

The gamma-ray analysis of PSR~J2302+4442 was similar to that of PSR~J2017+0603 (see Section \ref{J2017_gamma}). Figure \ref{phaso_J2302} shows light curves of PSR~J2302+4442 in radio and gamma rays. For events within 0.8$^\circ$ of the MSP the \emph{H}-test parameter is 415.8, also corresponding to a pulsation significance well above 10$\sigma$. The maximum of the radio profile at 1.4 GHz is at phase $\Phi_r = 0.960$, under the same convention for the absolute phasing as described in Section \ref{J2017_gamma}. We checked whether the structure between phase 0.25 and 0.4 comprises one or two gamma-ray peaks by plotting light curves with 10, 20, 30, 50 and 100 counts in each bin. We found that a sharp peak at phase $\sim$ 0.31 is clearly observed, whereas the possible component at phase $\sim$ 0.35 is not significant with the present dataset. We fitted the sharp structure at phase $\sim$ 0.31 as well as the second gamma-ray peak with Lorentzian functions above constant background. The peak positions and FWHM, as well as the radio-to-gamma-ray lag and gamma-ray peak separation are listed in Table \ref{params}. As for PSR~J2017+0603, the $\delta$ and $\Delta$ values follow the trend already noted by \citet{FermiPSRCatalog} for previously detected gamma-ray pulsars with known radio emission. However, we note in the case of PSR~J2302+4442 an alignment between the radio interpulse at phase $\sim$ 0.65 in Figure \ref{phaso_J2302}, and the second gamma-ray peak, indicating interesting frequency-dependence of the emission regions, if the radio and the gamma-ray emission features are indeed of common origin in the magnetosphere. 

\begin{figure}
\begin{center}
\epsscale{1.}
\plotone{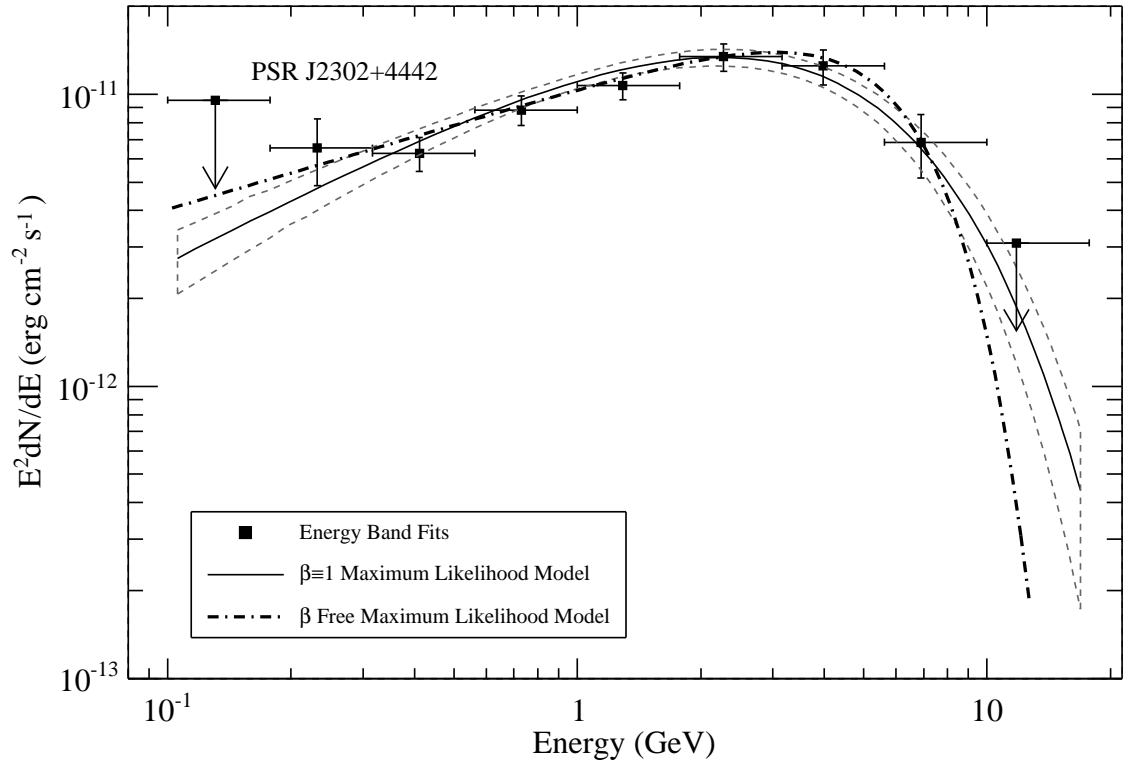}
\caption{Same as Figure \ref{spectre_J2017}, for PSR~J2302+4442. See text for details on the spectral analysis. \label{spectre_J2302}}
\end{center}
\end{figure}

The gamma-ray spectral parameters for PSR~J2302+4442 obtained from a fit with $\beta = 1$ are listed in Table \ref{params}, and Figure \ref{spectre_J2302} shows the corresponding energy spectrum. In this case, spectral parameters of sources within 6$^\circ$ from the pulsar were left free in the fit. The simple power-law model without cutoff is rejected at the 9$\sigma$ level. A spectral fit with the $\beta$ parameter in Equation (\ref{model}) left free gave $\beta = 2.4 \pm 0.7$. This value formally departs from the $\beta = 1$ assumption; however, we found that there is no statistical improvement of the fit compared to the simple exponentially cutoff power-law fit with the current data. As can be seen in Figure \ref{spectre_J2302}, the best-fit models with $\beta = 1$ and $\beta$ left free agree well except at the lowest and highest energies, where only upper limits could be measured. More data are thus needed to discriminate between the two models. Spectral parameters measured for $\beta = 1$ are again similar to those of gamma-ray MSPs observed so far \citep{Fermi8MSPs,FermiJ0034}. Finally, the energy flux listed in Table \ref{params} is consistent with that of the 1FGL Catalog source J2302.8+4443 measured above 0.1 GeV by \citet{Fermi1FGL} of (4.8 $\pm$ 0.4) $\times 10^{-11}$ erg cm$^{-2}$ s$^{-1}$. We therefore conclude that 1FGL~J2302.8+4443 is associated with the gamma-ray millisecond pulsar PSR~J2302+4442.

\section{Discussion}

\subsection{Gamma-ray light curve modeling}
\label{modeling}

Several of the MSPs detected by the \emph{Fermi} LAT in gamma rays have quite complex radio pulses, and PSRs~J2017+0603 and J2302+4442 are no exception. In contrast, their respective gamma-ray light curves are quite standard, exhibiting a familiar double-peak structure \citep{Fermi8MSPs,FermiPSRCatalog}. The gamma-ray and radio pulse shapes and relative lags motivated light curve modeling using standard outer magnetospheric pulsar models commonly employed to describe the light curves of younger pulsars and which have been successful in modeling earlier detected gamma-ray MSPs \citep{Venter2009}.  In such models the gamma-ray emission originates in gaps along the last open magnetic field lines, with emission from trailing field lines accumulating around a particular observer phase leading to intense peaks or ``caustics'', due to special relativistic effects \citep{Dyks2003}. In the outer gap (OG) model, two caustics originate from one magnetic pole \citep[e.g.,][]{Romani1995}, while caustics from both magnetic poles are visible in the case of the two-pole caustic (TPC) model. One may additionally consider a pair-starved polar cap (PSPC) model \citep{Muslimov2004,Muslimov2009,Harding2005} where the combination of perpendicular B-field strength and gamma-ray energies of the radiated photons are too low to lead to significant amounts of electron-positron pairs close to the stellar surface. In this case, the magnetosphere is ``pair-starved'' and no pair formation front is established, so that the primaries continue to accelerate along the B-field lines and emit curvature gamma-ray radiation up to near the light cylinder. The non-zero lags between the gamma-ray and radio pulses led us to model the radio using a phenomenological model proposed by \citet{Story2007}, where one assumes that the radio emission originates in a cone beam centered on the magnetic dipole axis at a single altitude.  Different combinations of inclination and observer angles $\alpha$ and $\zeta$ will result in zero, one or two radio peaks from each pole, along with different gamma-ray profile shapes, depending on how close an observer's line-of-sight sweeps with respect to the magnetic axis. 

We have used a Markov chain Monte Carlo (MCMC) maximum likelihood fitting technique to jointly model the gamma-ray and radio pulse profiles in order to statistically pick the best-fit emission model geometry \citep[details will be described in][]{Johnson2011}.  Additionally, we have generated simulations with 1$^{\circ}$ resolution in $\alpha$ as opposed to the 5$^{\circ}$ used in \citet{Venter2009} and included the Lorentz transformation of the magnetic field from the inertial observer's frame to the co-rotating frame which was missing in previous studies and advocated by \citet{Bai2010_2} as necessary for self-consistency. An MCMC technique involves taking random steps in parameter space, evaluating the likelihood at that step, and accepting the step based on the likelihood ratio with the previous step.  In particular, we use a Metropolis-Hastings method \citep{Hastings1970} to update the parameter state, accepting steps if the likelihood at the new step is greater than the previous step or if the ratio is greater than a random number $\in[0,1)$.  For each model fit we verify that our MCMC has converged using the method proposed by \citet{Gelman1992}.

The gamma-ray light curves are fit using Poisson likelihood and the radio profiles using a $\chi^{2}$ statistic.  In order to balance the contributions from the radio and gamma-ray data, and in particular to balance the high statistical precision of the radio data against our simple cone-beam model, we have used a relative error for the radio data equal to the average gamma-ray relative uncertainty in the on-peak region times the radio maximum. It is important to note that the choice of uncertainty for the radio profile can strongly affect the best-fit results. A smaller uncertainty will decrease the overall likelihood, which can in some cases lead to a different best-fit geometry favoring the radio light curve. For both MSPs we have taken the gamma-ray on-peak interval to be $\phi\in[0.25,0.75]$.  Our geometric models assume constant-emissivity gamma-ray emission extending from the stellar surface in the TPC model, while the minimum radius is set to the radius of the null charge surface (which depends on  magnetic azimuth and co-latitude) in the OG model.  For all simulations we have used a maximum emission altitude for the gamma rays of 1.2 R$_{\rm LC}$, where R$_{\rm LC} = c P / (2 \pi)$, with the added caveat that the emission not go beyond a cylindrical radius equal to 0.95 R$_{\rm LC}$.  We found that the likelihood surfaces are very multi-modal which can lead to a low acceptance rate and an incomplete exploration of the parameter space; therefore, we have implemented simulated tempering \citep{Marinari1992} with small-world chain steps \citep{Guan2006} in $\alpha$ and $\zeta$.  The MCMC parameter space includes $\alpha$, $\zeta$ (both with 1$^{\circ}$ resolution), gap width $w$ (with a resolution of 0.05, normalized to the polar cap radius), and phase-shift, which accounts for the fact that the definitions of phase zero are different between the data and our models.  Our MCMC is implemented in python using the \emph{scipy} module\footnote{See http://docs.scipy.org/doc/ for documentation} and the light curve fitting for each step is done using the \emph{scipy.optimize.fmin\_l\_fbgs\_b} multi-variate, bound optimizer \citep{Zhu1997}.

\begin{figure}
\begin{center}
\epsscale{1.}
\plotone{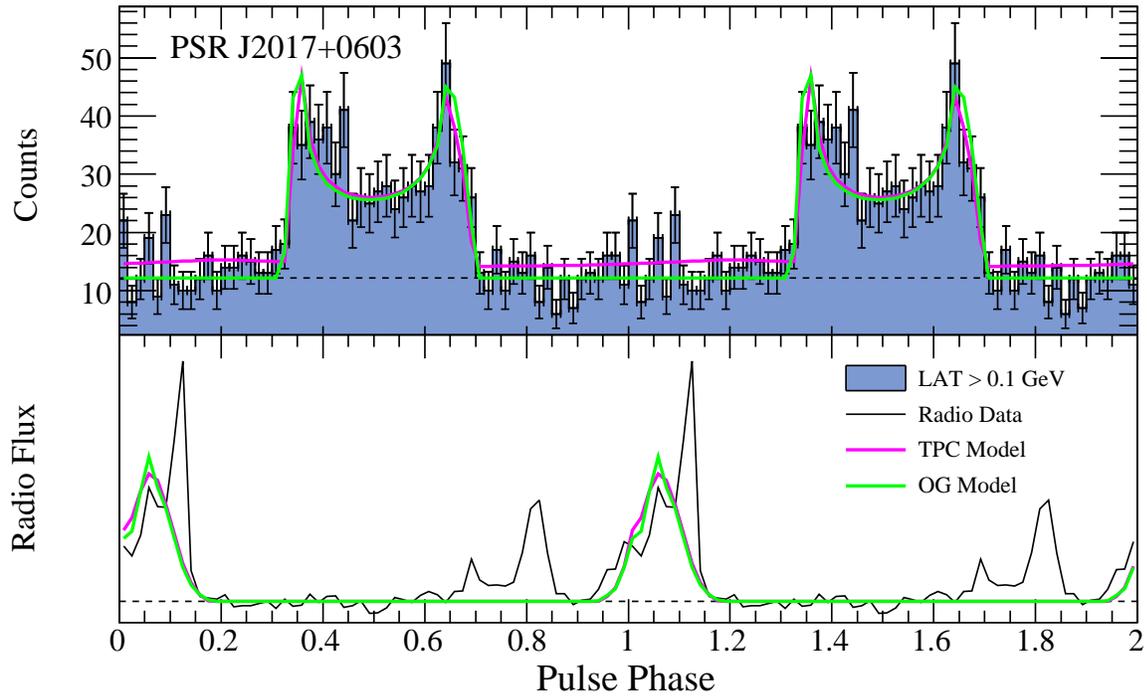}
\caption{Top: gamma-ray data and modeled light curves for PSR~J2017+0603 with 60 bins per rotation. Bottom: Nan\c cay 1.4 GHz radio profile and modeled light curves. Modeled light curves were made using $\alpha = 16^\circ$, $\zeta = 68^\circ$ and an infinitely thin gap for the TPC model, and $\alpha = 17^\circ$, $\zeta = 68^\circ$ and an infinitely thin gap for the OG geometry. See Section \ref{modeling} for emission altitude extents.\label{J2017_modeling}}
\end{center}
\end{figure}

\begin{figure}
\begin{center}
\epsscale{1.}
\plotone{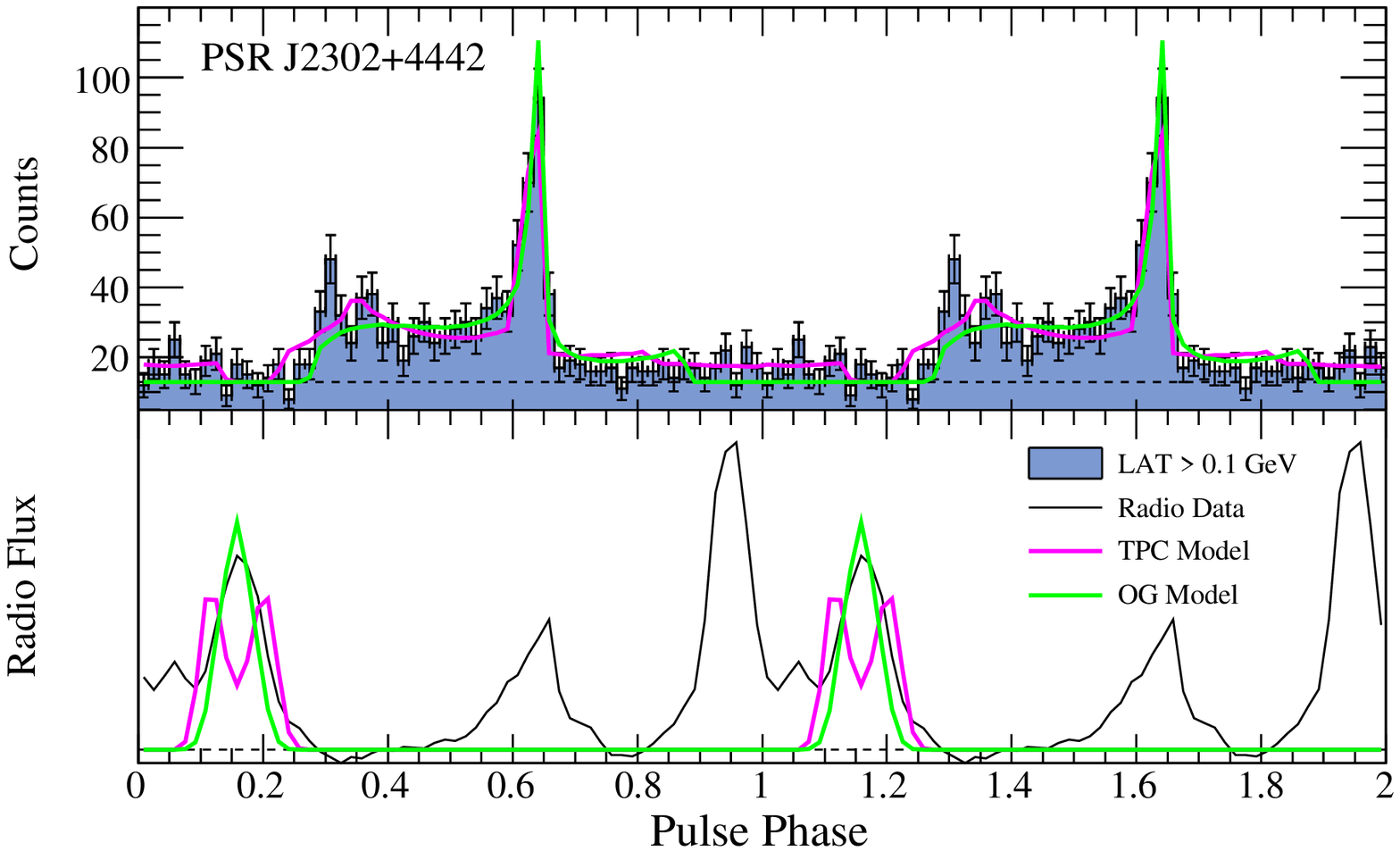}
\caption{Same as Figure \ref{J2017_modeling}, for PSR~J2302+4442. Modeled light curves were made using $\alpha = 58^\circ$, $\zeta = 46^\circ$ and an infinitely thin gap for the TPC emission geometry, and $\alpha = 63^\circ$, $\zeta = 39^\circ$ and an infinitely thin gap for the OG model. \label{J2302_modeling}}
\end{center}
\end{figure}

In order to match the data with our simulations we re-binned both the gamma-ray and radio data to 60 bins, see Figures \ref{J2017_modeling} and \ref{J2302_modeling}.  This has the effect of smoothing out very fine scale variations in the radio profile, but as we discuss below our radio profile simulations are not refined enough to reproduce these structures and thus fitting to the 60 bin radio profiles is sufficient to reproduce the general features, namely the gamma-to-radio lag.  For PSR~J2017+0603 we find best-fit solutions of $\alpha$ = 16$^{\circ}$ and $\zeta = 68^{\circ}$ with an infinitely thin gap for a TPC model and $\alpha$ = 17$^{\circ}$ and $\zeta = 68^{\circ}$ with an infinitely thin gap for an OG model. For PSR~J2302+4442 we find best-fit solutions of $\alpha$ = 58$^{\circ}$ and $\zeta = 46^{\circ}$ with infinitely thin gap for a TPC model and $\alpha$ = 63$^{\circ}$ and $\zeta = 39^{\circ}$ with infinitely thin gap for an OG model. When we find best-fit models with infinitely thin gap widths for both pulsars we do not think this represents the truth as a zero-width gap is unphysical; rather, we take this to mean that the best gap width is somewhere between 0 and 0.05 and the best-fit value of 0 is chosen only as a result of the resolution of our simulations. Note also that we have not yet calibrated the fitting procedure to address the significance of differences in $-$log(likelihood) so we cannot be more quantitative in discussing the preference of one model over another. However, for both MSPs differences in $-$log(likelihood) were close to 0, meaning that neither of TPC and OG geometries are preferred.

Neither of the model fits for PSR~J2017+0603 are able to produce a wide enough first gamma-ray peak but both produce the correct peak separation. Also, the model fits cannot reproduce all the features observed in the radio profile. However, the best-fit geometries are able to produce radio-to-gamma-ray lags close to what is observed. The situation is similar for PSR~J2302+4442, with both models matching the sharp second gamma-ray peak but neither is able to produce a strong enough first peak. The TPC model implies two small peaks near phase 0.3 for slightly different values of $\alpha$ and $\zeta$, close the to best-fit values. Tests have shown that lowering the maximum emission altitude can affect the prominence of these two peaks, which suggests that more investigation is merited in this parameter.  With more data the significance, or not, of this two-peaked structure will serve as a further discriminator between the models. Neither best-fit geometry produces produces two radio peaks with the correct spacing.  The TPC geometry does predict two closely spaced radio peaks while the OG geometry approximately matches the radio peak near 0.15 in phase.

For both MSPs, it is of interest to note that geometries with $\alpha$ and $\zeta$ both near 20$^{\circ}$ produce two radio peaks with approximately correct spacing but the resultant gamma-ray TPC light curves are similar to square waves while the gamma-ray emission in OG models is missed entirely. Clearly, our simple radio model does not adequately reproduce the data.  Both MSPs have at least three components in their radio profiles, while the model can only produce zero, one, or two peaks from each magnetic pole.  This points to more complex radio emission geometries, with radio emission from both magnetic poles visible, and likely that emission may occur higher up in the magnetosphere as has been suggested by \citet{Ravi2010}.

We also fit both MSPs with the PSPC model, though this is not as successful at producing sharp gamma-ray peaks. For both MSPs the fits predict $\alpha\sim70^{\circ}$ and $\zeta\sim80^{\circ}$ which suggest that we would see radio emission from both magnetic poles.  The gamma-ray PSPC models are able to reproduce the second, sharp peak for each MSP but have trouble matching the first peak properly.  The best-fit geometries result in more complex radio profiles but are still not able to match all of the observed features. For both MSPs the PSPC models are disfavored by the likelihood when compared to the TPC and OG fits.  Our modeling and fit results also show that there is still much to be learned about the radio beam structure.

\subsection{Gamma-ray efficiencies}
\label{efficiencies}

One can derive the total gamma-ray luminosity above 0.1 GeV and the efficiency of conversion of spin-down energy into gamma rays with the following expressions:

\begin{eqnarray}
\label{luminosity}
L_\gamma & = & 4 \pi f_\Omega G d^2,\\
\eta & = & L_\gamma / \dot E.
\end{eqnarray}

In these expressions, $d$ and $\dot E$ are the pulsar distance and spin-down energy, $f_\Omega$ is the correction factor depending on the viewing geometry defined above, and $G$ is the energy flux measured above 0.1 GeV. Table \ref{params} lists $L_\gamma$ and $\eta$ values under the assumption that $f_\Omega = 1$, and using the pulsar distances inferred from the NE2001 model (see Table \ref{ephem}). For both pulsars, gamma-ray efficiencies are found to be suspiciously large, and even greater than 100\% in the case of PSR~J2302+4442, which is unphysical. Overestimated $f_\Omega$ factors and distances are plausible explanations for the large efficiency values. The best-fit TPC and OG emission geometries discussed in Section \ref{modeling} predict geometrical correction factors of 0.48 and 0.30 for PSR~J2017+0603, leading to realistic gamma-ray efficiencies of 0.39 and 0.24, respectively. However, $f_\Omega$ factors calculated under TPC and OG geometries for PSR~J2302+4442 are 0.95 and 0.97, leading to gamma-ray efficiencies greater than 1.6. The distance inferred from the NE2001 model is therefore likely overestimated, or the model is incorrect. In addition, proper motions could make the apparent spin-down energy loss rates $\dot E$ larger than the intrinsic values because of the Shklovskii effect, thereby increasing gamma-ray efficiencies. The average efficiency of gamma-ray MSPs observed so far \citep{Fermi8MSPs,FermiJ0034} is $\sim$ 10\% (we excluded PSR J1614$-$2230, which also has an unphysical gamma-ray efficiency of 100\% with the NE2001 distance). Assuming an efficiency of 10\% for PSR~J2302+4442, we find that the distance has to be smaller by a factor of 4, which would place the pulsar at $d \lesssim 300$ pc. Note however that the X-ray energy flux $G_X$ of $\sim 3.1 \times 10^{-14}$ erg cm$^{-2}$ s$^{-1}$ measured between 0.5 and 3 keV leads to an X-ray efficiency of $4 \pi G_X d^2 / \dot E \sim 1.4 \times 10^{-3}$ if we assume the NE2001 distance of 1.18 kpc, while it decreases to $\sim 9 \times 10^{-5}$ with a distance of 300 pc. The former efficiency is very close to the $10^{-3}$ value empirically predicted by \citet{Becker1997} at these energies. The X-ray analysis therefore does not support such an important reduction of the distance. If the pulsar distance is indeed that small, a timing parallax $\pi = \frac{1}{d (\mathrm{kpc})} \gtrsim 3.3$ mas should be measurable with accumulated radio timing observations. This parallax could also be measured via the VLBI measurements being undertaken for all \emph{Fermi} pulsars\footnote{Cycle 3 \emph{Fermi} Guest Investigator proposal: S. Chatterjee et al.}.

\section{Conclusions}

In a search for radio pulsations at the position of \emph{Fermi} 1FGL catalog sources with the Nan\c cay radio telescope, we discovered two millisecond pulsars, PSRs~J2017+0603 and J2302+4442, both orbiting low-mass companion stars. Both pulsars were found to emit pulsed gamma-ray emission, indicating that they are associated with the previously unidentified gamma-ray sources. The gamma-ray light curves and spectral properties of the two MSPs are reminiscent of those of other gamma-ray MSPs observed previously.

Prior to \emph{Fermi}, error boxes of unidentified gamma-ray sources were much larger than radio telescope beams, making searches for pulsars difficult, as multiple pointings were required to cover the gamma-ray source contour entirely \citep[see for example][]{Champion2005}. Unassociated \emph{Fermi} LAT sources are typically localized to within 10 arcminutes, which is comparable to radio beam sizes and therefore makes radio pulsation searches easier and more efficient. With its improved localization accuracy and its homogeneous coverage of the gamma-ray sky, the \emph{Fermi} LAT is therefore revealing the population of energetic pulsars and millisecond pulsars, providing a complementary view of the Galactic population of pulsars, which has mostly been studied at radio wavelengths up to now.

\acknowledgments

The \emph{Fermi} LAT Collaboration acknowledges generous ongoing support from a number of agencies and institutes that have supported both the development and the operation of the LAT as well as scientific data analysis. These include the National Aeronautics and Space Administration and the Department of Energy in the United States, the Commissariat \`a l'Energie Atomique and the Centre National de la Recherche Scientifique / Institut National de Physique Nucl\'eaire et de Physique des Particules in France, the Agenzia Spaziale Italiana and the Istituto Nazionale di Fisica Nucleare in Italy, the Ministry of Education, Culture, Sports, Science and Technology (MEXT), High Energy Accelerator Research Organization (KEK) and Japan Aerospace Exploration Agency (JAXA) in Japan, and the K.~A.~Wallenberg Foundation, the Swedish Research Council and the Swedish National Space Board in Sweden.

Additional support for science analysis during the operations phase is gratefully acknowledged from the Istituto Nazionale di Astrofisica in Italy and the Centre National d'\'Etudes Spatiales in France.

The Nan\c cay Radio Observatory is operated by the Paris Observatory, associated with the French Centre National de la Recherche Scientifique (CNRS). The Green Bank Telescope is operated by the National Radio Astronomy Observatory, a facility of the National Science Foundation operated under cooperative agreement by Associated Universities, Inc. The Lovell Telescope is owned and operated by the University of Manchester as part of the Jodrell Bank Centre for Astrophysics with support from the Science and Technology Facilities Council of the United Kingdom.

The authors are greatly saddened by the passing of Professor Donald C. Backer in July 2010. He was not only an outstanding scientist and a leader of the instrumental developments leading to this paper, but he was also a wonderful friend.





\bibliographystyle{apj}

\bibliography{biblio}

\clearpage

\begin{table}
\begin{center}
\caption{Parameters for PSRs~J2017+0603 and J2302+4442. See Sections \ref{J2017_timing} and \ref{J2302_timing} for details on the measurement of these parameters. Numbers in parentheses are  the nominal 1$\sigma$ \textsc{tempo2} uncertainties in the least-significant digits quoted. \label{ephem}}
\begin{small}
\begin{tabular}{lcc}
\tableline\tableline
Parameter & PSR~J2017+0603 & PSR~J2302+4442 \\
\tableline
Right ascension (J2000) \dotfill & 20:17:22.7044(1) & 23:02:46.9796(7)\\
Declination (J2000)\dotfill & 06:03:05.569(4) & +44:42:22.090(5)\\
Rotational period, $P$ (ms)\dotfill  &  2.896215815562(2)   &  5.192324646411(7) \\
Period derivative, $\dot P$ (10$^{-21}$)\dotfill & 8.3(1) & 13.3(5) \\ 
Epoch of ephemeris, $T_0$ (MJD)\dotfill &  55000 &  55000  \\
Dispersion measure, DM (cm$^{-3}$ pc)\dotfill &  23.918(3)   &  13.762(6)  \\
Orbital period, $P_b$ (d)\dotfill &  2.198481129(6)   &  125.935292(3)  \\
Projected semi-major axis, $x$ (lt s)\dotfill &  2.1929239(7)   &  51.429942(3)  \\
Epoch of ascending node, $T_\mathrm{asc}$ (MJD)\dotfill &  55202.5321589(3)   &  55096.517187(3)  \\
$e \sin \omega$\dotfill & 0.0000023(6) & $-$0.00023537(6) \\
$e \cos \omega$\dotfill & $-$0.00000046(6) & $-$0.00044485(6) \\
Span of timing data (MJD)\dotfill &  54714 --- 55342   &  54712 --- 55342  \\
Number of TOAs\dotfill  &  71  &  130 \\
RMS of TOA residuals ($\mu$s)\dotfill &  3.23  &  6.46 \\
Units\dotfill & TDB & TDB \\
Solar system ephemeris model\dotfill  &  DE405  &  DE405 \\
Flux density at 1.4 GHz, $S_{1400}$ (mJy)\dotfill & 0.5(2) & 1.2(4) \\
\tableline
\multicolumn{3}{c}{Derived parameters} \\
\tableline
Orbital eccentricity, $e$\dotfill & 0.000005(2) & 0.0005033(2) \\
Mass function, $f$ (M$_\Sun$)\dotfill & 0.002342653(2) & 0.009209497(1)\\
Minimum companion mass, $m_{c}$ (M$_\Sun$)\dotfill & $\geq$ 0.18 & $\geq$ 0.30\\
Galactic longitude, $l$ ($^\circ$)\dotfill & 48.62 & 103.40\\
Galactic latitude, $b$ ($^\circ$)\dotfill & $-16.03$ & $-14.00$\\
Distance inferred from the NE2001 model, $d$ (kpc)\dotfill & 1.56 $\pm$ 0.16 & 1.18$^{+0.10}_{-0.23}$\\
Spin-down luminosity, $\dot E$ (10$^{33}$ erg s$^{-1}$)\dotfill & 13.43 & 3.74\\
Characteristic age, $\tau$ (10$^9$ yr)\dotfill & 5.55 & 6.20\\
Surface magnetic field strength, $B_\mathrm{s}$ (10$^8$ G)\dotfill & 1.57 & 2.66\\
Magnetic field strength at the light cylinder, $B_\mathrm{LC}$ (10$^4$ G)\dotfill & 5.86 & 1.73\\
\tableline
\end{tabular}\\
\end{small}
\end{center}
\end{table}


\begin{table}
\begin{center}
\caption{Light curve and spectral parameters of PSRs~J2017+0603 and J2302+4442 in gamma rays, fixing $\beta = 1$ in Equation (\ref{model}). See Sections \ref{J2017_gamma} and \ref{J2302_gamma} for details on the measurement of these parameters. Peak positions, widths and separations are given in phase units, between 0 and 1.\label{params}}
\begin{small}
\begin{tabular}{lcc}
\tableline\tableline
Parameter & PSR~J2017+0603 & PSR~J2302+4442 \\
\tableline
First peak position, $\Phi_1$ & 0.348 $\pm$ 0.009 & 0.310 $\pm$ 0.021 \\
First peak full width at half-maximum, FWHM$_1$ & 0.248 $\pm$ 0.054 & 0.033 $\pm$ 0.013 \\
Second peak position, $\Phi_2$ & 0.636 $\pm$ 0.005 & 0.629 $\pm$ 0.003 \\
Second peak full width at half-maximum, FWHM$_2$ & 0.050 $\pm$ 0.013 & 0.037 $\pm$ 0.006 \\
Radio-to-gamma-ray lag, $\delta$ & 0.225 $\pm$ 0.009 $\pm$ 0.002 & 0.350 $\pm$ 0.021 $\pm$ 0.002 \\
Gamma-ray peak separation, $\Delta$ & 0.288 $\pm$ 0.010 & 0.320 $\pm$ 0.021 \\
\tableline
Spectral index, $\Gamma$ & 1.00 $\pm$ 0.16 $\pm$ 0.16 & 1.25 $\pm$ 0.13 $\pm$ 0.14 \\
Cutoff energy, E$_c$ (GeV) & 3.12 $\pm$ 0.57 $\pm$ 0.75 & 2.97 $\pm$ 0.51 $\pm$ 0.54 \\
Photon flux, $F$ ($> 0.1$ GeV) (10$^{-8}$ cm$^{-2}$ s$^{-1}$) & 2.21 $\pm$ 0.31 $\pm$ 0.11 & 3.34 $\pm$ 0.38 $\pm$ 0.20 \\
Energy flux, $G$ ($> 0.1$ GeV) (10$^{-11}$ erg cm$^{-2}$ s$^{-1}$) & 3.71 $\pm$ 0.24 $\pm$ 0.19 & 3.94 $\pm$ 0.22 $\pm$ 0.10 \\
Luminosity, L$_\gamma$ / f$_\Omega$ (10$^{33}$ erg s$^{-1}$) & 10.79 $\pm$ 1.72 $\pm$ 1.66 & 6.57 $^{+0.87}_{-1.85}$ $^{+0.80}_{-1.82}$\\
Efficiency, $\eta$ / f$_\Omega$ & 0.80 $\pm$ 0.13 $\pm$ 0.12 & 1.75 $^{+0.23}_{-0.49}$ $^{+0.21}_{-0.48}$ \\
\tableline
\end{tabular}\\
\end{small}
\end{center}
\end{table}

\end{document}